\documentclass[english,aps,prb,twocolumn]{revtex4}
\usepackage{amsmath}
\usepackage{amssymb}
\usepackage{psfrag}
\usepackage{graphicx}
\usepackage{epsfig}
\usepackage{epstopdf}
\usepackage{fontenc}

\makeatletter
\usepackage{babel}
\makeatother
\begin{document} 

\preprint{This line only printed with preprint option}

\title{Exchange-controlled single-electron-spin rotations in quantum dots}

\author{W. A. Coish and Daniel Loss}
\affiliation{Department of Physics and Astronomy, University of Basel, Klingelbergstrasse
82, 4056 Basel, Switzerland}


\begin{abstract}
  We show theoretically that arbitrary coherent rotations can be
  performed quickly (with a gating time $\sim 1\,\mathrm{ns}$) and with
  high fidelity on the spin of a single confined electron using
  control of exchange only, without the need for spin-orbit coupling
  or ac fields. We expect that implementations of this scheme would achieve gate
  error rates on the order of $\eta\lesssim 10^{-3}$ in GaAs quantum dots,
  within reach of several known error-correction protocols.
\end{abstract}
\maketitle

The elementary building-blocks for universal quantum computing are a
two-qubit entangling operation, such as the $\mbox{\sc cnot}$-gate or
$\sqrt{\mbox{\sc swap}}$-gate and arbitrary single-qubit rotations.
For qubits based on single electron spins confined to quantum dots,
\cite{loss:1998a} recent experiments have now achieved the two-qubit
$\sqrt{\mbox{\sc swap}}$-gate \cite{petta:2005a} and single-spin
coherent rotations.\cite{koppens:2006a}
If these operations are to be used in a viable quantum information
processor, they must be performed with a sufficiently small gate error
per operation $\eta\ll 1$.  The threshold values of $\eta$ required for
effective quantum error correction depend somewhat on error models and
the particular error-correction protocol, but current estimates are in
the range $\eta<10^{-2}-10^{-4}$.\cite{steane:2003a, knill:2005a} 
To achieve these low error rates, new schemes must be developed to
perform quantum gates quickly and accurately
within the relevant coherence times.

Previous proposals \cite{engel:2001a} and recent implementations
\cite{koppens:2006a} for single-spin rotation have relied on ac
magnetic fields to perform electron-spin resonance (ESR). In ESR,
difficulties with high-power ac fields limit
single-spin Rabi frequencies to values that are much smaller than the
operation rates typically associated with two-qubit gates
mediated by exchange.\cite{petta:2005a} To circumvent these problems while 
still achieving fast coherent
single-qubit rotations, there have been several proposals to use
exchange or electric-field (rather than magnetic-field) control 
of electron spin states.  These proposals aim to perform rotations 
on multiple-spin
encoded qubits,\cite{hanson:2006a,kyriakidis:2006a} or require strong
spin-orbit interaction,
\cite{rashba:2003a,stepanenko:2004a,flindt:2006a,golovach:2006a}
coupling to excited orbital states,\cite{tokura:2006a} or rapid
pulsing of magnetic fields.\cite{wu:2004a} Qubits encoded
in two states having different orbital wave functions 
are susceptible to dephasing
through fluctuations in the electric environment, even in the idle
state.\cite{coish:2005a,hu:2006a,stopa:2006a}  Proposals that make
use of the spin-orbit interaction
\cite{rashba:2003a,stepanenko:2004a,flindt:2006a,golovach:2006a} are
restricted to systems where the spin-orbit coupling is sufficiently
strong, excluding promising architectures such as quantum dots
made from Si:SiGe \cite{friesen:2003a} and carbon nanotubes or
graphene sheets.\cite{mason:2004a,graeber:2006a,trauzettel:2006a} Sufficiently rapid pulsing of
magnetic fields \cite{wu:2004a} may not be feasible in GaAs, where
the electron-spin coherence time is on the order of $\tau_c\sim
10\,\mathrm{ns}$.\cite{koppens:2005a,petta:2005a}

\begin{figure}[t]
\scalebox{0.6}{\includegraphics{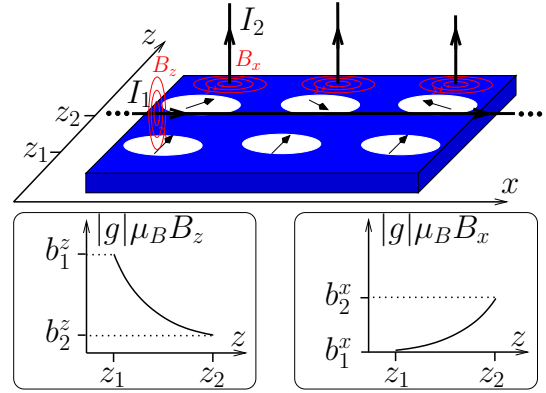}}
\caption{\label{fig:setup} (Color online) Possible setup to implement
the scheme proposed here.  Ancillary electron spins at $z_1$ are
maintained in a polarized state with a large Zeeman field $b_1^z$
along $z$.  Qubit spins at $z_2$ are free to precess in a weaker
effective Zeeman field lying in the $x$-$z$ plane:
$\pmb{\Delta}=(b_2^x,0,b_2^z-J/2)$. Here, $J$ is the exchange coupling
between the qubit and ancillary spins and $\mathbf{b}_2$ is the qubit
Zeeman field in the absence of exchange. When $b_1^z\gg b_2^z \gg b_2^x$,
$z$-rotations are performed if $J\approx 0$ and $x$-rotations are achieved
when $J\approx 2 b_2^z$.}
\end{figure}Here we propose to perform single-qubit rotations
in a way that would marry the benefits of demonstrated fast electrical control of
the exchange interaction \cite{petta:2005a} with the benefits of
naturally long-lived single-electron spin qubits.\cite{loss:1998a}
Our proposal would operate in the absence of spin-orbit coupling and
would act on single electron spins without the use of ac
electromagnetic fields, in the presence of a fixed Zeeman field 
configuration (Fig. \ref{fig:setup}).  This scheme applies
to confined electrons in any structure with a locally
controllable potential.  Specifically, this scheme may be applied to 
electrons above liquid helium, bound to gated phosphorus
donors in silicon, and in quantum dots formed in a GaAs 
two-dimensional electron gas, nanowires, carbon nanotubes, or graphene.
  
\begin{figure}
\scalebox{0.9}{\includegraphics{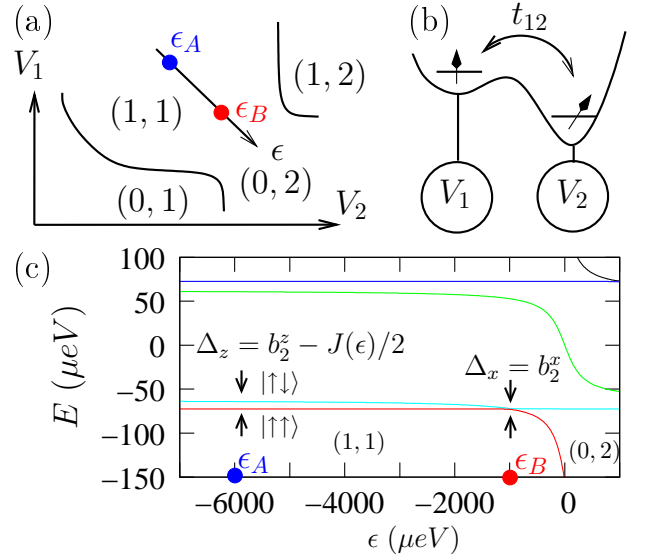}}
\caption{\label{fig:scheme} (Color online) (a) Charge stability
diagram indicating the ground-state charge configuration $(N_1,N_2)$
for local dot potentials $V_1,\,V_2$.  In the $(1,1)$ configuration,
the exchange interaction $J(\epsilon)$ can be tuned by shifting the
double-dot potential difference $\epsilon\sim V_2-V_1$. (b) When the electron
spin in dot 1 is polarized, the qubit electron acquires a Zeeman shift
given by $t_{12}^2/\epsilon=-J(\epsilon)/2$ due to virtual hopping processes that
are allowed for spin-down, but forbidden for spin-up due to the Pauli
principle. (c) Energy spectrum of the Hamiltonian given in
Eq. (\ref{eq:Hamiltonian}) in the presence of a strong inhomogeneous
magnetic field.  
}
\end{figure}

 
\emph{Hamiltonian}--We begin from a standard tunneling model for the
two lowest orbital levels of a double quantum dot, including tunnel
coupling $t_{12}$, on-site repulsion $U_\mathrm{c}$, nearest-neighbor
repulsion $U^\prime_\mathrm{c}$, local electrostatic potentials $V_{1(2)}$
and a local Zeeman field $\mathbf{b}_{1(2)}$ on dot 1(2) (see
Refs. [\onlinecite{vanderwiel:2003a,coish:2006a}] and references
therein):
\begin{multline}
\mathcal{H}=-\sum_{l\sigma}V_{l}n_{l\sigma}+U_\mathrm{c}\sum_{l}n_{l\uparrow}n_{l\downarrow}+U^{\prime}_\mathrm{c}\prod_{l}(n_{l\uparrow}+n_{l\downarrow})\\
+t_{12}\sum_{\sigma}\left(d_{1\sigma}^{\dagger}d_{2\sigma}+d_{2\sigma}^{\dagger}d_{1\sigma}\right)-\sum_{l}\mathbf{S}_{l}\cdot\mathbf{b}_{l}.\label{eq:Hamiltonian}\end{multline}
Here we have set $\hbar=1$, $d_{l\sigma}$ annihilates an electron in dot
$l=1,2$ with spin $\sigma$, $n_{l\sigma}=d_{l\sigma}^\dagger d_{l\sigma}$ is the usual number
operator, and 
$\mathbf{S}_l=\frac{1}{2}\sum_{s,s^\prime}c^\dagger_{ls}\pmb{\sigma}_{s,s^\prime}c_{ls^\prime}$ 
is the spin density on dot $l$.  We
choose $|\epsilon \pm \delta b^z| \gg |t_{12}|$, $|\delta b^z| \gtrsim |t_{12}|$, with $\delta
b^z=(b_1^z-b_2^z)/2$ and $\epsilon=V_2-V_1-U_\mathrm{c}+U^\prime_\mathrm{c}$,
which favors the $(1,1)$ charge state (where
$(N_1,N_2)$ denotes a state with $N_{1(2)}$ electrons on dot 1(2), see
Fig. \ref{fig:scheme}).  Additionally, we require a large Zeeman
field along $z$ in dot 1 ($|b_1^{z}|\gg|b_1^{x,y}|$) so that the spin on
dot 1 is frozen into its spin-up ground state.  For simplicity, we furthermore
choose $b_2^y=0$. Eq. (\ref{eq:Hamiltonian}) then reduces to the
following low-energy effective Hamiltonian for the spin on dot 2:
\begin{equation}
\mathcal{H}_\mathrm{eff}=-\frac{1}{2}\pmb{\Delta}\cdot\pmb{\sigma};\;
\pmb{\Delta}=(b_2^x,0,b_2^z-J(\epsilon)/2).
\label{eq:Heff}
\end{equation}

When $|\epsilon| \gg |\delta b^z|$,
$J(\epsilon)\approx-2t_{12}^2/\epsilon$. Thus, for a fixed Zeeman field $\mathbf{b}_2$,
the direction and magnitude of the effective field $\pmb{\Delta}$ can be
tuned with gate voltages via its dependence on $\epsilon$ (see
Fig. \ref{fig:scheme}(c)).  Eq. (\ref{eq:Heff}) follows
directly from a much more general Hamiltonian of the form
$H=J(\epsilon)\mathbf{S}_1\cdot\mathbf{S}_2-\sum_l \mathbf{b}_l\cdot\mathbf{S}_l$ in the
limit where $|\mathbf{b}_1|\gg|\mathbf{b}_2|,J$, and so this scheme is
not limited to the particular Hamiltonian given in
Eq. (\ref{eq:Hamiltonian}), which neglects the
long-ranged nature of the Coulomb interaction and excited orbital
states.  The long-ranged part of the Coulomb interaction (the exchange
integral) contributes a small fraction to $J(\epsilon)$ compared to the
tunneling contribution when the out-of-plane magnetic field is zero
\cite{burkard:1999a} and contributions to $J(\epsilon)$ due to excited
orbital states \cite{stopa:2006a} are a small correction when
$|\epsilon|<J_{(0,2)}$, where $J_{(0,2)}$ is the single-dot exchange coupling
on dot 2.  Outside of this range of validity, the functional 
form $J(\epsilon)$ could be obtained empirically, as has been done 
in Ref. [\onlinecite{laird:2006a}].


\emph{Qubit gates}--Arbitrary single-qubit rotations can
be achieved with the appropriate composition of the Hadamard gate
($H$) and $\pi/8$ gate ($T$) \cite{nielsen:2000a}:
\begin{equation} H=\frac{1}{\sqrt{2}}\left(\begin{array}{cc} 1 & 1\\ 1
& -1\end{array}\right),\;\;T=\left(\begin{array}{cc} 1 & 0\\ 0 &
e^{i\pi/4}\end{array}\right).
\label{eq:gates}\end{equation} Up to a global phase, $T$  corresponds to a rotation about
$z$ by an angle $\phi=\pi/4$. This operation can be performed with high
fidelity by allowing the qubit spin to precess coherently for a switching time $t_\mathrm{s} = \phi/\Delta_z$
at the operating point $\epsilon_A$ in Fig. \ref{fig:scheme}(a), where
$\Delta_z\gg \Delta_x$. The $H$ gate can be implemented by pulsing $\epsilon$ (see
Fig. \ref{fig:scheme}(c)) from $\epsilon_A$, where $\Delta_z\approx b_2^z$ to $\epsilon_B = -
t_{12}^2/b_2^z$, where $\Delta_z\approx 0$ and back. The pulse is achieved with a
characteristic rise/fall time $\tau$, and returns to $\epsilon=\epsilon_A$ after spending
the pulse time $t_\mathrm{p}$ at $\epsilon=\epsilon_B$.  If $b_2^x\ll b_2^z$, 
$\mathcal{H}_\mathrm{eff}$ induces approximate $z$-rotations during 
the rise/fall time, and $x$-rotations when $\epsilon=\epsilon_\mathrm{B}$. The 
entire switching process (with total switching 
time $t_\mathrm{s}=t_\mathrm{p}+4\tau$) is 
described by a time evolution operator
$\mathcal{U}=\mathcal{T}e^{i\int_0^{t_\mathrm{s}}dt\pmb{\Delta}(t)\cdot\pmb{\sigma}/2}$,
which, for $b_2^x\ll b_2^z$, is thus approximately given by
\begin{equation} \mathcal{U}\approx\mathcal{U}(\phi_x,\phi_z) =
R_{\hat{z}}\left(-\frac{\phi_z}{2}\right) R_{\hat{x}}\left(-\phi_x\right)
R_{\hat{z}}\left(-\frac{\phi_z}{2}\right),\label{eq:UApprox}
\end{equation} where $\phi_x = \Delta_x t_\mathrm{p}$ and
$\phi_z=\int_0^{t_\mathrm{s}}dt\Delta_z(t)$. Here, $R_{\hat{n}}(\phi)$ is a rotation
about the $\hat{n}$-axis by angle $\phi$. When $\phi_x=\pi/2$ and $\phi_z=\pi$,
Eq. (\ref{eq:UApprox}) gives an $H$ gate, up to a global phase:
$\mathcal{U}(\pi/2,\pi)= iH$.

\emph{Errors}--We quantify gate errors with the error rate
$\eta=1-\mathcal{F}$, where $\mathcal{F}$ is the average gate fidelity,
defined by
\begin{equation} \mathcal{F}=\frac{1}{4\pi}\int d\Omega
\overline{\mathrm{Tr}\left(U\rho_\mathrm{in}(\theta,\phi)U^\dagger\tilde{U}\rho_\mathrm{in}(\theta,\phi)\tilde{U}^\dagger\right)}.\label{eq:Fidelity}
\end{equation} Here,
$\rho_\mathrm{in}(\theta,\phi)=\left|\theta,\phi\right>\left<\theta,\phi\right|$, where
$\left|\theta,\phi\right>=\cos(\theta/2)\left|\uparrow\right>+e^{i\phi}\sin(\theta/2)\left|\downarrow\right>$
indicates an inital spin-1/2 coherent state in the qubit basis (the
two-dimensional Hilbert space spanned by the (1,1) charge state and
spin-up on dot 1),
$U=H$ or $T$ is the ideal intended single-qubit gate operation, and
$\tilde{U}=\mathcal{T}\exp{\left[-i\int_0^{t_\mathrm{s}}dt\mathcal{H}(t)\right]}$
is the true time evolution of the system under the time-dependent
Hamiltonian $\mathcal{H}(t)$.  The overbar indicates a
Gaussian average over fluctuations in the classical Zeeman field
$\mathbf{b}_{2}$, which reproduces the effects of
hyperfine-induced decoherence due to an unknown static nuclear
field when $\sigma\ll \left|\overline{\mathbf{b}_2}\right|$ \cite{coish:2004a}:
\begin{equation} \overline{f(\mathbf{b}_2)}=\int
\frac{d^3b_2}{\left(2\pi\sigma\right)^{3/2}}
\exp\left(\frac{\left(\mathbf{b}_2-\overline{\mathbf{b}_2}\right)^2}{2\sigma^2}\right)f(\mathbf{b}_2).
\label{eq:average}\end{equation} 
For a gated lateral GaAs quantum dot,
$\sigma_N^2=\overline{\left(\mathbf{b}_2-\overline{\mathbf{b}_2}\right)^2}=3\sigma^2$
due to hyperfine fluctuations has been measured, giving $\sigma_N=0.03\,\mu
eV$.\cite{koppens:2005a}

\begin{figure}
\scalebox{1.2}{\includegraphics{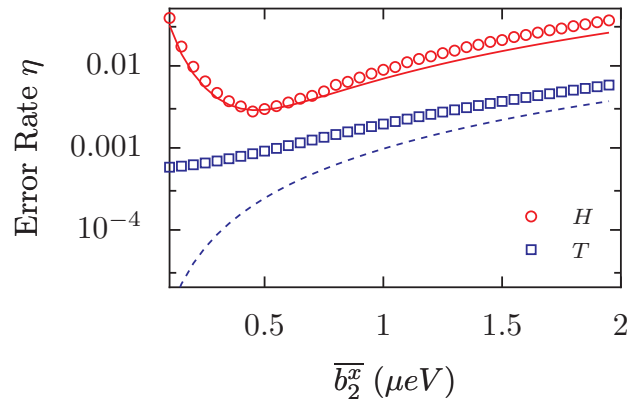}}
\caption{\label{fig:errorrates} (Color online) Error rates for $H$ and
  $T$ gates.  For these plots we
have chosen the parameters $t_{12}=100\,\mu eV,\;b_1^z=135\,\mu eV$,
 $\overline{b_2^z}=10\,\mu eV,\;\overline{b_2^x}=1\,\mu eV$, and
$\epsilon_A=-6\,\mathrm{m}eV$. For the Hadamard gate, these values result in
$\epsilon_B = -t_{12}^2/\overline{b_2^z} = -1000\,\mu eV$, a pulse time $t_\mathrm{p}=\pi\hbar/2
\overline{b_2^x}=1\,\mathrm{ns}$, and a rise/fall time $\tau \approx
\pi\hbar/4\overline{b_2^z}=50\,\mathrm{ps}$. Symbols give the results of numerical
integration of the time-dependent Schr\"odinger equation for the
Hadamard gate ($H$, circles) and $\pi/8$ gate ($T$, squares).  Lines 
give the estimates for gating error from Eq. (\ref{eq:GateErrorRates}).}
\end{figure}

Based on the above protocol for gating operations, and assuming a
coherence time $\tau_c$ for the qubit spins, a suitable parameter 
regime for high-fidelity single-qubit operations is given by the following hierarchy:
\begin{equation} 1/\tau_c\ll \overline{b_2^x}\ll \overline{b_2^z}\ll t_{12}\lesssim b_1^z \ll |\epsilon_B| < |\epsilon_A|.
\label{eq:validity}\end{equation}
The first inequality in Eq. (\ref{eq:validity}) guarantees that 
$x$-rotations are achieved with high fidelity at the operating 
point $\epsilon=\epsilon_B$.  The second inequality allows for high-fidelity 
$z$-rotations at $\epsilon=\epsilon_A$. The third and fourth inequalities are
required to ensure that $\overline{b_2^z}$ can be cancelled by exchange 
$J\simeq 2t_{12}^2/\epsilon$, and the last two inequalities guarantee 
that the population of (0,2) (the double occupancy $D\simeq \left(t_{12}/\epsilon\right)^2$) remains
small, which limits errors due to leakage and orbital dephasing (see
below). When $\tau_c$ is dominated by
hyperfine fluctuations, $1/\tau_c\sim \sigma_N$.  In this case, we give a
set of values for these parameters satisfying Eq. (\ref{eq:validity})
in the caption of Fig. \ref{fig:errorrates}. The effective
Zeeman-field gradient given here could be achieved under the following
circumstances: (a) a GaAs double quantum dot with the nuclei in dot 1 at
near full polarization, which would produce a maximum effective 
Zeeman splitting of $b_1^z \simeq 135\,\mu eV$ (high polarizations could be
achieved, e.g., through optical pumping \cite{imamoglu:2003a} or
transport \cite{rudner:2006a}), or (b) a nanomagnet
neighboring a carbon nanotube or graphene
double quantum dot with $g$-factor $g=2$ and inter-dot separation 
$\Delta L\sim 1\,\mu \mathrm{m}$ or an InAs nanowire double quantum 
dot with g-factor $g=8$ and inter-dot separation 
$\Delta L\sim 100\,\mathrm{nm}$, either of which would require a
magnetic-field gradient on the order of $\Delta B/\Delta L\sim 1\mathrm{T}/\mu \mathrm{m}$.
Comparable field gradients have already been achieved
experimentally.\cite{wrobel:2004a}  Alternatively, the ancillary spins could be
polarized with the exchange field from a neighboring ferromagnet,
high $g$-factor material, or stripline currents 
(see Fig. \ref{fig:setup}).  The values we have used for the detuning
parameter $\epsilon$ and tunnel coupling $t_{12}$ are of the same
order as those used in previous experiments. \cite{petta:2005a} 

Within the validity of the two-dimensional effective Hamiltonian 
$\tilde{U}\approx\exp{\left\{-i\mathcal{H}_\mathrm{eff}(\epsilon)t_\mathrm{s}\right\}}$, it is
straightforward but tedious to calculate rotation errors at
$\epsilon=\epsilon_\mathrm{A}$ ($z$-rotations: $U=R_{\hat{z}}(\phi)$) and $\epsilon=\epsilon_\mathrm{B}$
($x$-rotations: $U=R_{\hat{x}}(\phi)$) using the expressions in
Eqs. (\ref{eq:Fidelity},\ref{eq:average}).\footnote{The
  integral over initial states in Eq. (5) can 
  be replaced by a discrite sum, which presents a significant 
  simplification (see M.~D.~Bowdrey, D.~K.~L.~Oi, A.~J.~Short, 
  K.~Banaszek, and J.~A.~Jones, Phys. Lett. A \textbf{294}, 258 (2002))}
The error rate for $z$-rotations is dominated by the
misalignment of the average field $\overline{\mathbf{b}_2}$ with the
$z$-axis and is thus small in the ratio
$\overline{b_2^x}/\overline{b_2^z}$.  For a rotation by 
angle $-\phi$ (to leading order in $\overline{b_2^x}/\overline{b_2^z}$),
this error rate is
\begin{equation}
\label{eq:Zroterror}
\eta^z(\phi)\approx\frac{2}{3}\left(\frac{\overline{b_2^x}}{\overline{b_2^z}}\right)^2\sin^2\left(\frac{\phi}{2}\right).
\end{equation} When $\overline{\Delta_z}=0$ (at $\epsilon=\epsilon_\mathrm{B}$), the error rate for $x$-rotations is
dominated by hyperfine fluctuations, and is therefore small in
$\sigma_N/\overline{b_2^x}$.  We find this error rate for an
$x$-rotation by angle $-\phi$ (to leading order in $\sigma_N/\overline{b_2^x}$) is
\begin{equation}
\label{eq:Xroterror}
\eta^x(\phi)\approx\left(\frac{\phi^2}{18}+\frac{4}{9}\sin^2\left(\frac{\phi}{2}\right)\right)\left(\frac{\sigma_N}{\overline{b_2^x}}\right)^2.
\end{equation} We estimate the error in $T$ using $\eta^z(\phi)$
with $\phi=\pi/4$.  To estimate the error in $H$, we use 
Eq. (\ref{eq:UApprox}) in combination with
Eqs. (\ref{eq:Zroterror}) and (\ref{eq:Xroterror}), assuming the
errors incurred by each rotation are independent.  These estimates
give
\begin{equation} \eta \approx
\left\{\begin{array}{cc}
\eta^z\left(\frac{\pi}{4}\right), & (U=T),\\
\eta^x\left(\frac{\pi}{2}\right)+2\eta^z\left(\frac{\pi}{2}\right), & (U=H).
\end{array}\right.
\label{eq:GateErrorRates}\end{equation} From
Eq. (\ref{eq:GateErrorRates}) we find the error rate for $H$
is always larger than that for $T$ and reaches a minimum 
at an optimal value of $\overline{b_2^x}$.  The optimal value
of $\overline{b_2^x}$ and $\eta$ at this point are:
\begin{equation} b_2^{x,\mathrm{opt}} = \sqrt{C|\overline{b_2^z}|\sigma_N},\;\;
\eta(b_2^{x,\mathrm{opt}}) = \frac{4}{3}C\frac{\sigma_N}{|\overline{b_2^z}|},
\end{equation} where $C$ is a numerical prefactor $C=\sqrt{1/3+\pi^2/48}
\simeq 0.73$.  Using the measured value $\sigma_N=0.03\,\mu eV$ and $b_2^z=10\,\mu
eV$, we find an optimized error rate of $\eta\sim 10^{-3}$.  Here we have
included the most dominant error mechanisms.  There are many other
potential sources of error, which we discuss in the following.  All 
numerical estimates are based on the parameter values
given in the caption of Fig \ref{fig:errorrates}.

The error due to leakage to the (0,2) singlet state or misalignment of
$\mathbf{b}_1$ due to the hyperfine interaction in leading-order
perturbation theory is given by
$\sim\max\left[\left(\sigma_N/b_1^z\right)^2,\left(t_{12}/\epsilon_A\right)^2\right]\sim
10^{-4}$.  

If switching is done too slowly during the Hadamard gate,
the qubit states will follow the adiabatic eigenbasis, introducing an
additional source of error.  We estimate this error to be $1-P\approx\alpha$,
where $P=e^{-\alpha}$ is the Landau-Zeener tunneling probability, determined by
\cite{zener:1932a}
\begin{equation} \alpha = \frac{\pi |\overline{b_2^x}|^2}{|dJ(t)/dt|}\approx
\frac{\pi|\overline{b_2^x}|^2\epsilon_B^2\tau}{2 t_{12}^2|\Delta\epsilon|}\sim 10^{-4}.
\end{equation} Here, we have used $dJ(t)/dt\approx-2\dot{\epsilon}t_{12}^2/\epsilon_B^2$,
with $|\dot{\epsilon}|\approx|\Delta\epsilon|/\tau$, where $\Delta\epsilon=\epsilon_A-\epsilon_B$.  In the opposite limit,
$\alpha\gg 1$, the qubit spin could be read out via charge
measurements \cite{petta:2005a} by
sweeping slowly to large positive $\epsilon$, where the qubit state
$\left|\uparrow\right>$ would be adiabatically converted to the (0,2)
ground-state singlet, or initialized by sweeping in the opposite
direction (see Fig. \ref{fig:scheme}(c)).


In systems with finite spin-orbit coupling, the transverse-spin 
decay time $T_2$ is limited by the
energy relaxation time $T_1$ (i.e., $T_2=2T_1$ \cite{golovach:2004a}),
so it is sufficient to analyze this error in terms of $T_1$. $T_1$ in
quantum dots can now be measured,\cite{elzerman:2004a} giving $T_1\sim
1\,\mathrm{ms}$ at fields of $B\approx6\mathrm{T}$ ($g^*\mu_B B\simeq 135\,\mu eV$).
\cite{amasha:2006a} This value gives an error estimate on the order
of $t_\mathrm{s}/T_1\lesssim 10^{-6}$ for a switching time $t_\mathrm{s}\simeq
1\,\mathrm{ns}$.  

Finally, rapid voltage-controlled gating in this
scheme is made possible only because the electron spin states are
associated with different orbital wave functions during pulsing, which
also makes these states susceptible to orbital dephasing.  The
associated dephasing time is, however, strongly suppressed in the
limit where the double occupancy is small: $D\approx (t_{12}/\epsilon)^2\ll 1$.  In
particular, the dephasing time for the two-electron system is
$\tau_\phi^{(2)}\gtrsim D^{-2}\tau_\phi^{(1)}$, \cite{coish:2005a} where
$\tau_\phi^{(1)}\approx1\,\mathrm{ns}$ \cite{hayashi:2003a} is the single-electron
dephasing time in a double quantum dot.  This gives an error estimate
of $t_\mathrm{s}/\tau_\phi^{(2)}\lesssim 10^{-4}$, using $t_\mathrm{s}\approx
1\,\mathrm{ns}$ and $D\sim10^{-2}$ at the operating point $\epsilon=\epsilon_B$.  It
should be possible to further suppress orbital
dephasing by choosing the operating point $\epsilon_B$ to coincide with a
``sweet spot'', where $dJ(\epsilon_B)/d\epsilon=0$.
\cite{coish:2005a,hu:2006a,stopa:2006a}

\emph{Numerical analysis}--To confirm the validity of the
approximations made here and to verify the smallness of error
mechanisms associated with leakage and finite pulse times, we have
numerically integrated the time-dependent Schr\"odinger equation for the
Hamiltonian given in Eq. (\ref{eq:Hamiltonian}) in the basis of the
(0,2) singlet state and four (1,1) states (including spin).  We have
used the pulse scheme described following Eq. (\ref{eq:gates}) and evaluated
the gate error rates for $T$ and $H$ from the fidelity in
Eq. (\ref{eq:Fidelity}).  For the Hadamard gate we used the symmetric
pulse shape
\begin{equation} \epsilon(t) = \left\{\begin{array}{c}
\epsilon_0+\frac{\Delta\epsilon}{2}\tanh\left(\frac{2\left[t-2\tau\right]}{\tau}\right),
0<t<\frac{t_\mathrm{s}}{2}\\
\epsilon_0+\frac{\Delta\epsilon}{2}\tanh\left(\frac{2\left[t_\mathrm{s}-2\tau-t\right]}{\tau}\right),
\frac{t_\mathrm{s}}{2}<t<t_\mathrm{s}
                                                       \end{array}\right.,
\label{eq:HadamardPulse}
\end{equation} where $\epsilon_0=\left(\epsilon_A+\epsilon_B\right)/2$ and $\Delta\epsilon = \epsilon_B-\epsilon_A$.
The pulse time $t_\mathrm{p}$ and rise/fall time
$\tau=(t_\mathrm{s}-t_\mathrm{p})/4$ were fixed using
\begin{equation} t_\mathrm{p}=\frac{\pi}{2
b_2^x},\;\;\pi=\int_0^{t_\mathrm{s}} \Delta_z(t)dt,
\label{eq:tpandtau}
\end{equation} where the solution to the above integral equation was
found numerically.  The results of our numerics are shown in
Fig. \ref{fig:errorrates}.  To implement the integral (Eq. (\ref{eq:average}))
numerically, we have performed a Monte
Carlo average over 100 Overhauser fields, sampled from a uniform
Gaussian distribution using the experimental value $\sigma_N=0.03\,\mu eV$.
Error bars due to the finite sample of Overhauser fields are smaller 
than the symbol size. We find good agreement between the analytical
and predicted error rate for $T$ in the limit of large
$\overline{b_2^x}$ (the saturation value for $\eta$ at low $\overline{b_2^x}$ is consistent
with our estimates of $\sim10^{-4}$ for error due to leakage).
Additionally, we find reasonable agreement with our estimate for the
$H$-gate error rate, confirming that we have identified the dominant
error mechanisms. This gives us confidence that an error rate on the
order of $\sim 10^{-3}$ should be achievable with this proposed scheme.

\emph{Acknowledgments}--We thank G. Burkard, H.-A. Engel, M. Friesen,
V. N. Golovach, J. Lehmann, D. Lidar, B. Trauzettel, and
L. M. K. Vandersypen for useful discussions.  We
acknowledge financial support from JST ICORP, EU NoE MAGMANet, the
NCCR Nanoscience, and the Swiss NSF.

\bibliographystyle{apsrev} 
\bibliography{qubitrot}

\end{document}